\documentclass[10pt]{article}
\usepackage{graphicx}

\usepackage{caption}
\usepackage{subcaption}

\def\met{\mbox{${\hbox{$E$\kern-0.65em\lower-.1ex\hbox{\bf{/}}}}_T$}} 
\def\mex{\mbox{${\hbox{$E$\kern-0.65em\lower-.1ex\hbox{\bf{/}}}}_x$}~} 
\def\mey{\mbox{${\hbox{$E$\kern-0.65em\lower-.1ex\hbox{\bf{/}}}}_y$}~} 

\def\Title#1{\begin{center} {\Large #1 } \end{center}}
\def\Author#1{\begin{center}{ \sc #1} \end{center}}
\def\Address#1{\begin{center}{ \it #1} \end{center}}

\newcommand\pubblock{\rightline{\begin{tabular}{l} Proceedings of the Second Annual LHCP\\ \pubnumber\\
         \pubdate  \end{tabular}}}

\newenvironment{Abstract}{\begin{quotation} \begin{center} 
             \large ABSTRACT \end{center}\bigskip 
      \begin{center}\begin{large}}{\end{large}\end{center} \end{quotation}}

\newenvironment{Presented}{\begin{quotation} \begin{center} 
             PRESENTED AT\end{center}\bigskip 
      \begin{center}\begin{large}}{\end{large}\end{center} \end{quotation}}

\def\beq{\begin{equation}}
\def\eeq#1{\label{#1}\end{equation}}
\def\eeqn{\end{equation}}

\def\beqa{\begin{eqnarray}}
\def\eeqa#1{\label{#1}\end{eqnarray}}
\def\eeqan{\end{eqnarray}}

\let\bar=\overbar

\def\Dslash{\not{\hbox{\kern-4pt $D$}}}
\def\dslash{\not{\hbox{\kern-2pt $\del$}}}

\def\msb{{\bar{\ssstyle M \kern -1pt S}}}

\usepackage{color}

\newcommand\pubnumber{ CMS-PAS EXO-12-047 }

\newcommand\pubdate{\today}

\def\affiliation{
On behalf of the CMS Experiment, \\
Department of Physics \\
Brown University, Providence, RI 02912, U.S.A }

\begin{document}

\large
\begin{titlepage}
\pubblock

\vfill
\Title{  Shining Light On Dark Matter with the CMS Experiment }
\vfill

\Author{ Zeynep Demiragli}
\Address{\affiliation}
\vfill
\begin{Abstract}

We present a search for large extra dimensions and dark matter pair-production using events with a photon and missing transverse energy in pp collisions at $\sqrt{s} =8$ TeV.  This search is done with the data taken by the CMS experiment at the LHC corresponding to an integrated luminosity of 19.6 fb$^{-1}$. We find no deviations with respect to the standard model expectation and improve the current limits on several models.

\end{Abstract}
\vfill

\begin{Presented}
The Second Annual Conference\\
 on Large Hadron Collider Physics \\
Columbia University, New York, U.S.A \\ 
June 2-7, 2014
\end{Presented}
\vfill
\end{titlepage}
\def\thefootnote{\fnsymbol{footnote}}
\setcounter{footnote}{0}
%

\normalsize 


\section{Introduction}

Events from $pp$ collisions at the Large Hadron Collider (LHC) containing a photon ($\gamma$) of large transverse momentum ($p_T$) and missing transverse energy ($\met$) are used to investigate two extensions to the standard model (SM). One involves a model of dark matter (DM), which is now widely accepted as the dominant contribution to the matter density of the universe.  At the LHC, DM can be directly produced in the reaction $qq \rightarrow \gamma \chi \bar{\chi}$, where the photon is radiated by one of the incoming quarks and is used to trigger on the event. 

The $\gamma + \met$ final state is also sensitive to models involving extra spatial dimensions. The Arkani-Hamed, Dimopoulos, and Dvali (ADD) model provides a possible solution to the disparity between two seemingly unrelated fundamental scales of nature:  the electroweak unification scale ($M_{EW} \approx$ 100 GeV) and the Planck scale ($M_{P} \approx 10^{19} $ GeV). The process $qq \rightarrow \gamma$ G, where the graviton G escapes detection, further motivates the search for events with a single high-$p_T$ photon and $\met$.

In the ADD framework, the brane may also fluctuate in the extra dimensions. If the brane is assumed to be flexible, the fluctuations of the brane could give rise to scalar particles called Branons. These particles could be pair produced in association with SM particles at the LHC.

The primary background for the $\gamma + \met$ signal is the irreducible SM background from $Z (\rightarrow \nu \bar{\nu})\gamma~$ production. This and other SM backgrounds such as $W\gamma$, $W (\rightarrow l \nu)~$, $Z (\rightarrow l \bar{l})\gamma~$, QCD multijet,  $\gamma +$ jet and $\gamma\gamma$ events and background from beam halo are accounted for in the analysis.

\section{Selection and Observation}

Events containing an isolated photon with transverse energy larger than 145 GeV and $\met$ larger than 140 GeV are selected from a data sample corresponding to an integrated luminosity of 19.6 fb$^{-1}$ using single $\gamma$ and $\gamma + \met$ triggers. A topological cut of $\Delta \phi (\met - \gamma) > 2$ is imposed and events with significant hadronic or leptonic activity are rejected. In order to reduce the contamination of events with non intrinsic $\met$, a $\chi^2$ function is constructed and minimized using the unclustered energy ($\mathcal{U} = - (\met + \sum{p_{T}})$) in the event. 
\begin{equation}\label{eq:1a}
   \chi^2 = \sum_{i=objects} \left( \frac{p_{T}^{reco} - {p_{T}}}{\sigma_{p_{T}}} \right)_i^2 + \left(  \frac{\mathcal{U}} {\sigma_{\mathcal{U}}} \right)^2   
\end{equation}
For events with intrinsic $\met$, the minimization will result in large $\chi^2$ values. Therefore, events with small $\chi^2$ probability are rejected. 

Backgrounds that originate out of time with the nominal collision event are estimated from data by examining the transverse energy distribution of the electromagnetic (EM) cluster and the arrival time of the signal. Events with electrons misidentified as photons arise mainly from $W (\rightarrow e \nu)$ events and are estimated with $Z (\rightarrow e e)$ events in data. Events with jets misidentified as photons are estimated using a control sample of EM-enriched QCD events. Backgrounds from $Z (\rightarrow \nu \bar{\nu})\gamma$, $W\gamma$, $Z (\rightarrow l \bar{l})\gamma$, $\gamma +$ jet and $\gamma\gamma$ events are estimated from simulation.

\begin{table}[htb]
\begin{center}
\begin{tabular}{ccc}  
Process &  $Z\gamma$ Control Region Estimate & $W\gamma$ Control Region Estimate \\ \hline
 $Z(\rightarrow \nu \bar{\nu})\gamma$  &   57.9  $\pm$ 6.3     &   0.4 $\pm$ 0.1 \\
  $W(\rightarrow l \nu)\gamma$  &   37.6 $\pm$ 6.0    &   99.3 $\pm$ 22.2 \\
 $W \rightarrow e \nu $  &   15.1$\pm$ 1.5    &   1.2 $\pm$ 0.2  \\
 Jet $\rightarrow \gamma$ Fakes  &   9.7  $\pm$ 3.2    &   17.0  $\pm$ 5.3  \\ 
 Other Backgrounds  &   4.5 $\pm$ 1.2    &   8.12 $\pm$ 0.5  \\  \hline
 Total Background & 124.8 $\pm$ 12.8 & 126.1 $\pm$ 22.8 \\ 
 Oberved Data  & 123 & 104 \\ \hline
\end{tabular}
\caption{  Expected yields compared to observed data in the $Z\gamma$ and $W\gamma$ control regions. }
\label{tab:table1}
\end{center}
\end{table}

To check that the SM background predictions properly model the data, control regions with small potential signal contamination are established. A first region dominated by the $Z\gamma$ process is obtained by inverting the topological cut between the $\gamma$ and $\met$. A second region dominated by the $W\gamma$ process is obtained by inverting the lepton veto. The observed data in both control regions is well described by the background predictions as shown in Table \ref{tab:table1}.

\begin{figure}[htb]
\centering
\begin{subfigure}{0.48\textwidth}
\centering
\includegraphics[height=2.8in]{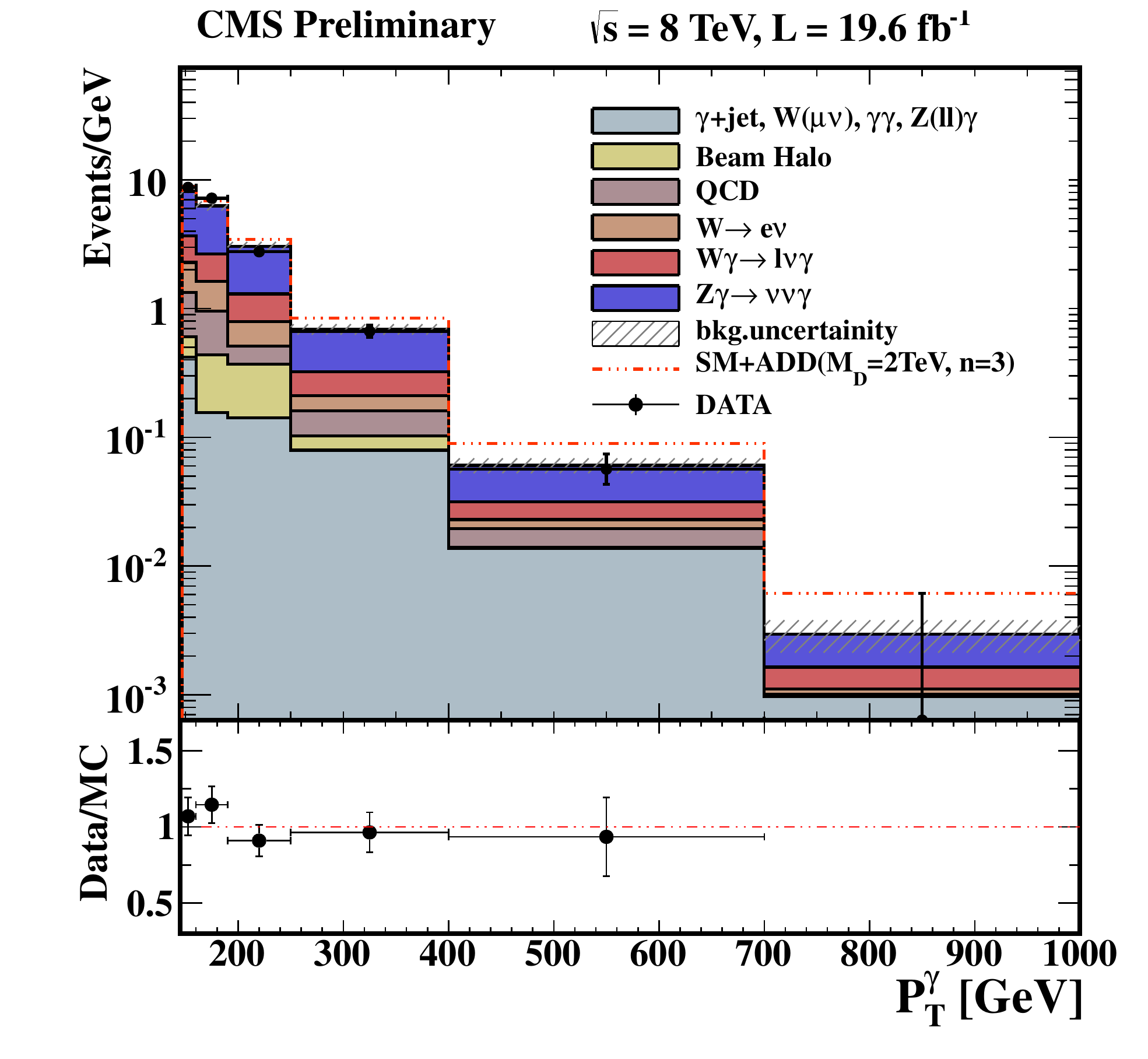}
\caption{}
\end{subfigure}
$\;\;\;\;$
\begin{subfigure}{0.48\textwidth}
\centering
\includegraphics[height=2.8in]{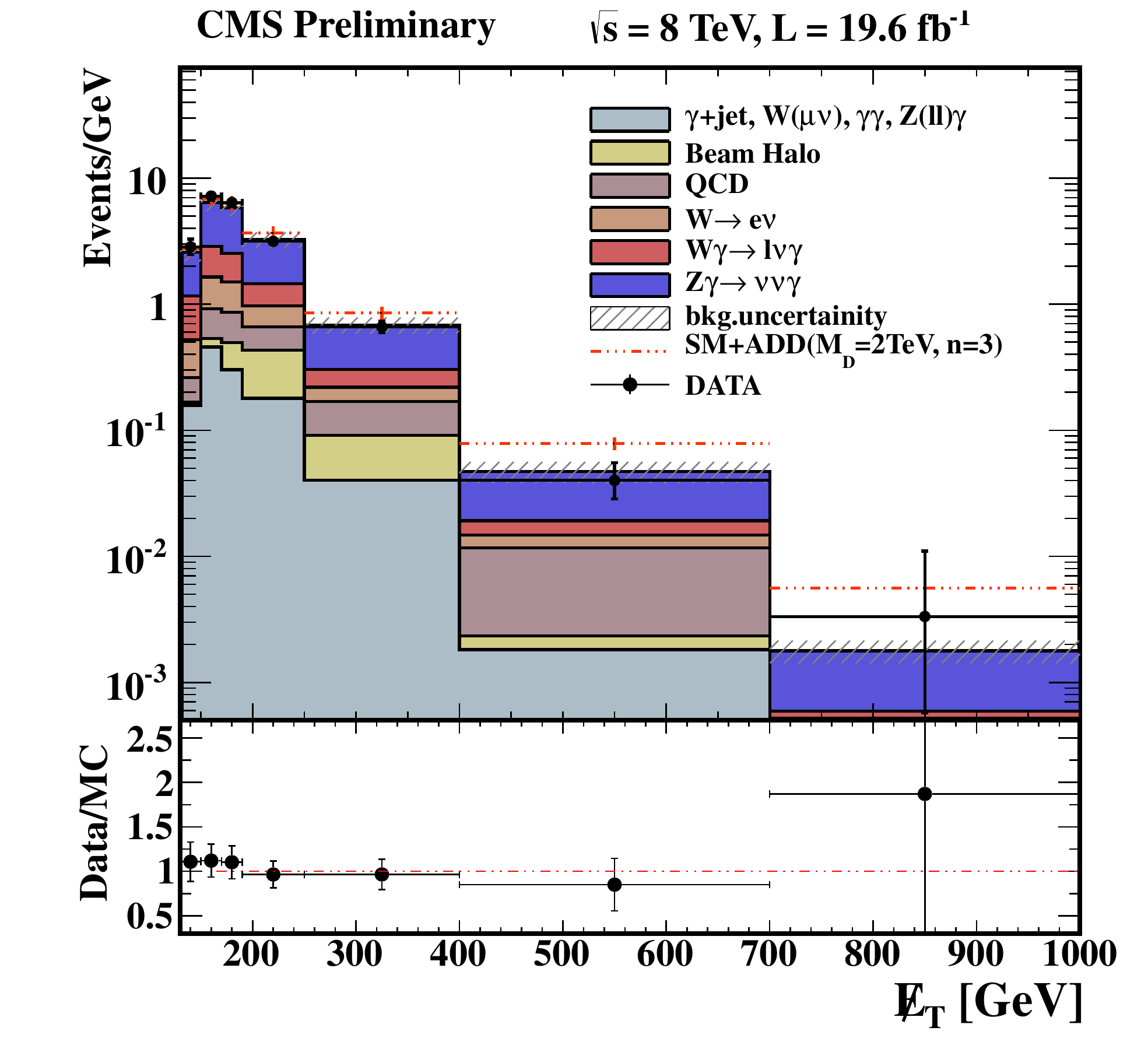}
\caption{}
\end{subfigure}
\caption{The photon $p_T$ and $\met$ distribution for the candidate sample, compared with the estimated contributions from SM backgrounds. Also shown is the prediction from the ADD model with $M_D$ = 2 TeV and n = 3. The background uncertainty represents the combined statistical and systematic uncertainty.} \label{fig:result}
\end{figure}

In the candidate data sample, the 630 observed events agree within uncertainty with the total expected background of 612.6 $\pm$ 50.0 events. Distributions of the photon $p_T$ and $\met$ for the selected candidate events along with the estimated backgrounds are shown in Fig. \ref{fig:result}.

\section{Interpretations}
\begin{figure}[htb]
\centering
\begin{subfigure}{0.48\textwidth}
\centering
\includegraphics[height=2.4in]{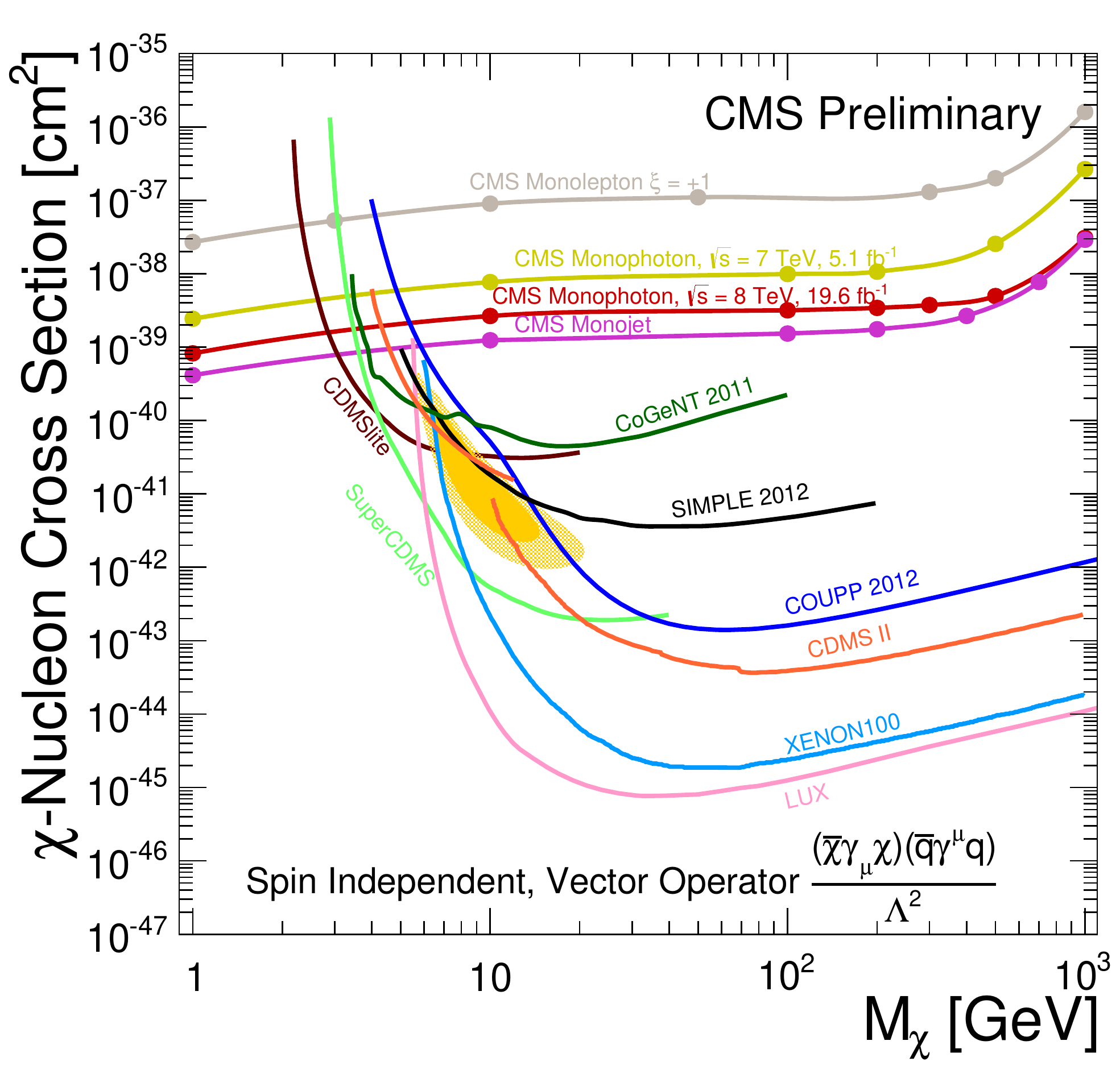}
\caption{}
\end{subfigure}
\begin{subfigure}{0.48\textwidth}
\centering
\includegraphics[height=2.4in]{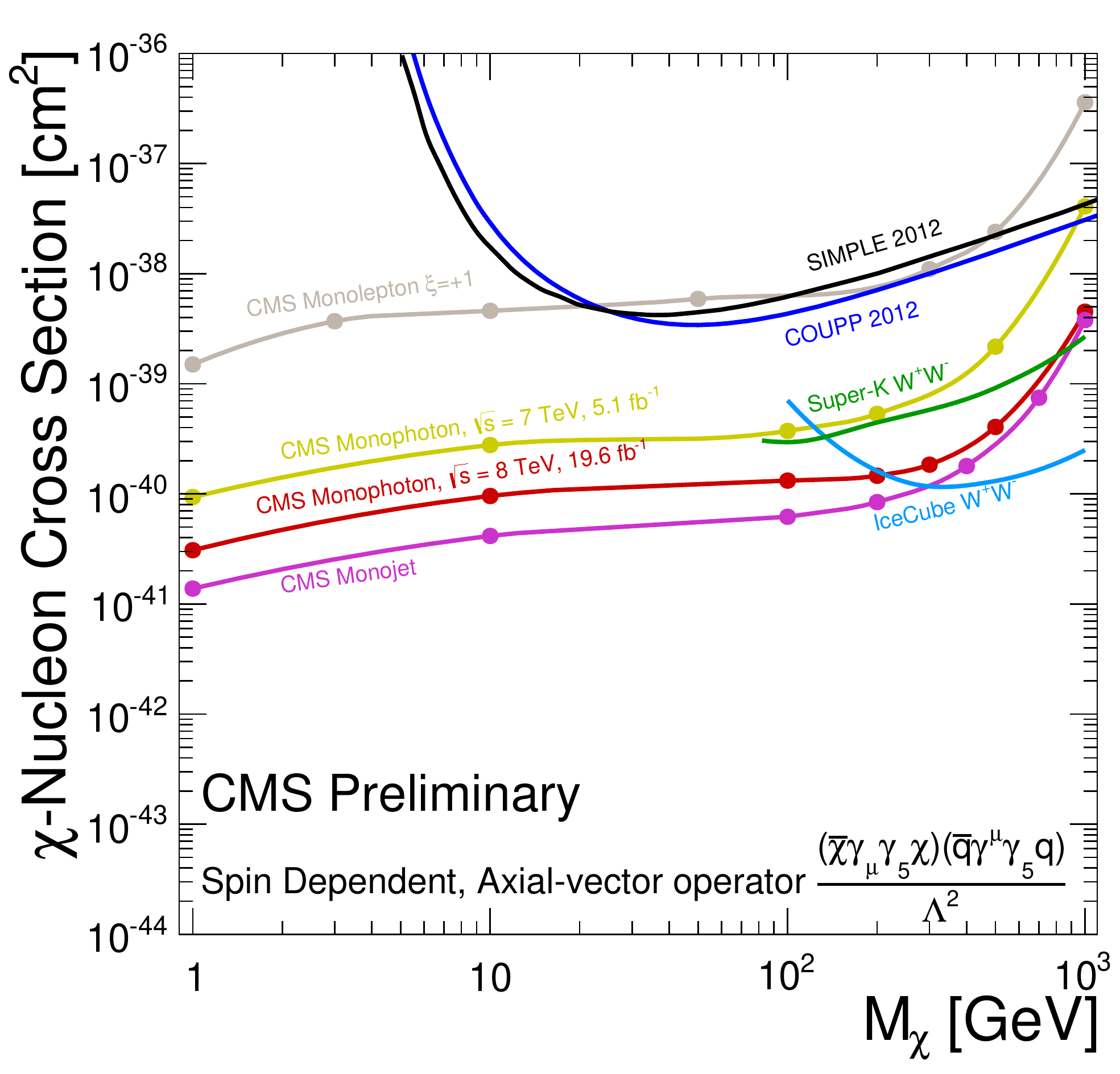}
\caption{}
\end{subfigure}
\\
\hspace{3pt}
\begin{subfigure}{0.48\textwidth}
\centering
\includegraphics[width=2.5in]{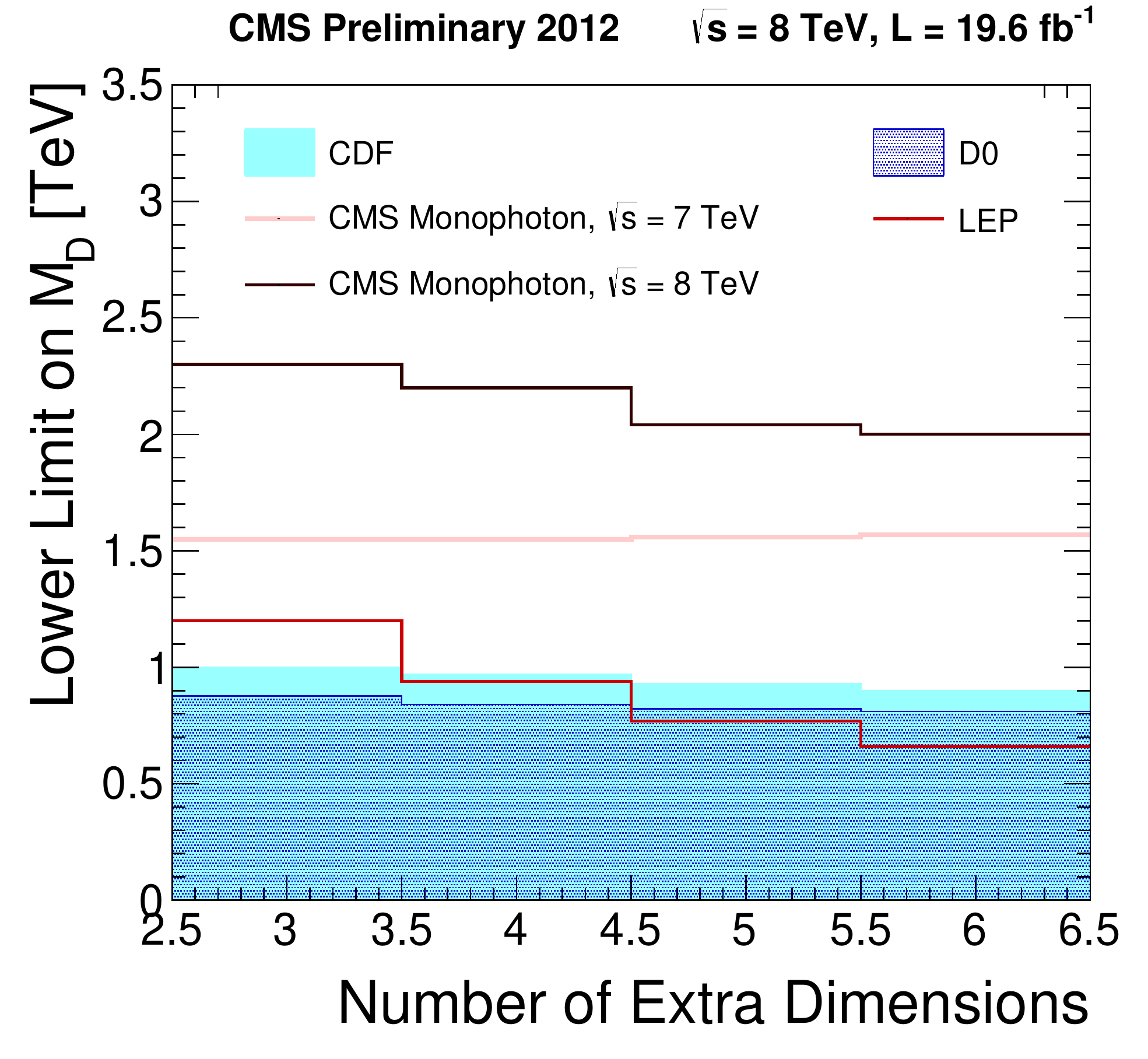}
\caption{}
\end{subfigure}
\hspace{6pt}
\begin{subfigure}{0.48\textwidth}
\centering
\includegraphics[width=2.6in, height=2.3in]{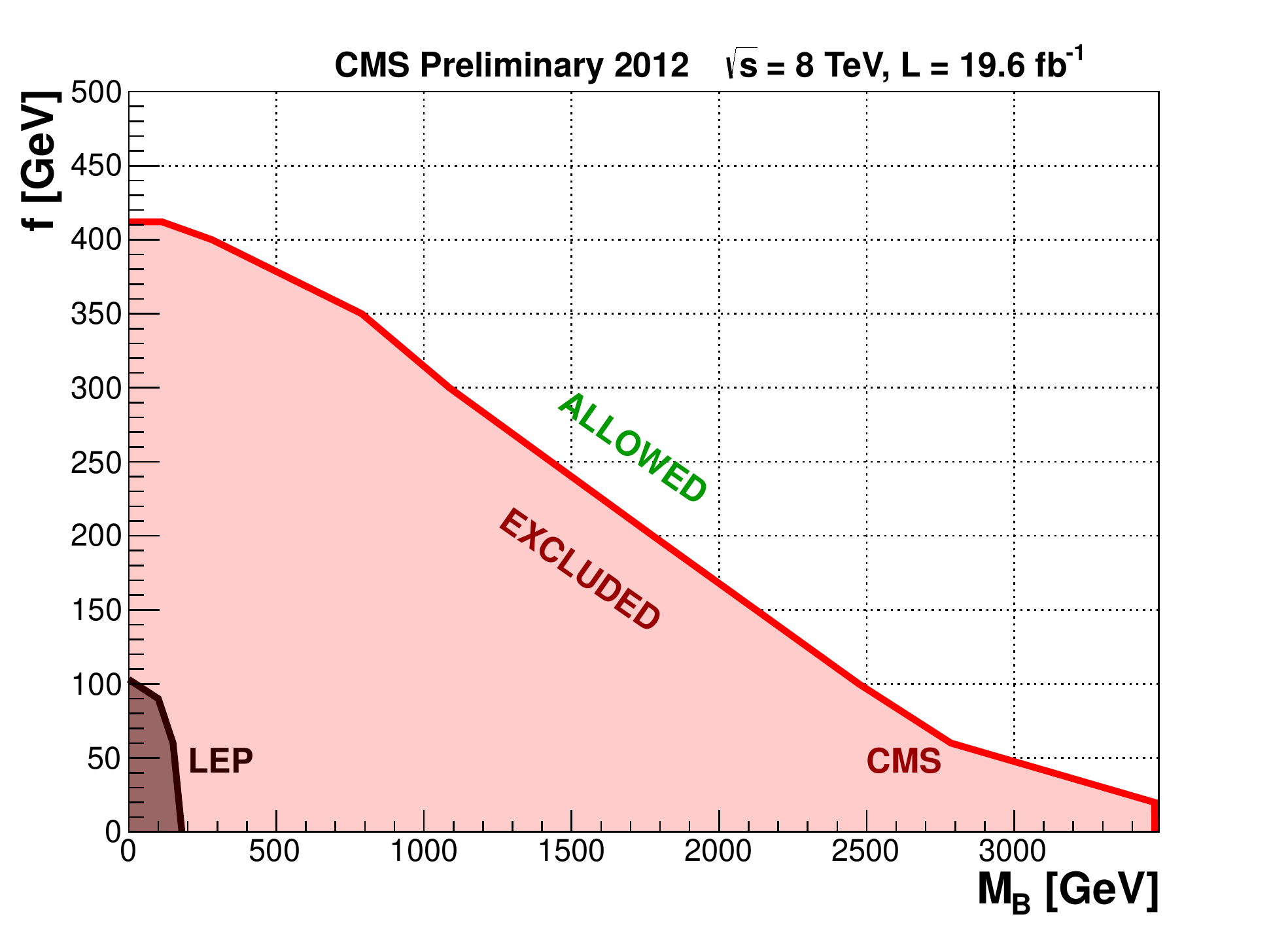}
\caption{}
\end{subfigure}
\caption{The 90\% CL upper limits on the $\chi$-nucleon cross section as a function of $M_\chi$ for (a) spin-independent and (b) spin-dependent scattering. Also shown are the limits from selected experiments with published results. Limits on $M_D$ as a function of n (c) compared to results from similar searches at the Tevatron and LEP.  Limits on $f$ as a function of $M_B$ (d) compared to similar searches at LEP.} \label{fig:figuredm}
\end{figure}

90\% CL upper limits are placed on the DM production cross section as a function of $M_\chi$ assuming either vector or axial-vector operators. The cross section limits are converted into lower limits on the cutoff scale $\Lambda$, which are then translated into upper limits on the $\chi$-nucleon cross sections calculated within the effective field theory framework. These limits can then be compared to direct and indirect detection dark matter experiments. The CMS limits are the most stringent over a wide range of DM mass hypotheses in the case of spin dependent couplings as seen in Fig. \ref{fig:figuredm}.


95\% CL upper limits are placed on the production cross section in the ADD model and are translated into exclusions on the model parameters $M_D$ and n. Masses $M_D < 2.30$ TeV are excluded at 95\% CL for n = 3. For massless branons, the brane tension $f$ is found to be greater than 412 GeV and branon masses $M_B < 3.5$ TeV are excluded at 95\% CL for low brane tension (20 GeV). These are the most stringent bounds to date as seen in Fig. \ref{fig:figuredm}.

\section{Conclusions}

In summary, the analysis of $\gamma + \met$ production in pp collisions at $\sqrt{s}$ = 8 TeV was used to derive stringent upper limits on the vector and axial-vector contributions to the $\chi$-nucleon scattering cross section. This search is complementary to searches for elastic $\chi$-nucleon scattering or $\chi\chi$ annihilation. In addition the analysis attained the most stringent limits to date on branon masses and on an effective extra-dimensional Planck scale from the $\gamma + \met$ channel.


\end{document}